%% file: paper.tex
\documentclass[10pt, letterpaper, twocolumn]{article} 

\input{structure.tex} 

\usepackage{url}
\usepackage{appendix}

\usepackage{booktabs}   
\usepackage{subcaption} 

\usepackage{color}
\usepackage{xcolor}
\usepackage{listings}
\usepackage{mdframed} 
\usepackage{float}
\usepackage{dblfloatfix}
\usepackage{wrapfig}

\usepackage[font=scriptsize]{caption}
\definecolor{antiquewhite}{rgb}{0.98, 0.92, 0.84}
\definecolor{azure(web)(azuremist)}{rgb}{0.94, 1.0, 1.0}
\definecolor{cosmiclatte}{rgb}{1.0, 0.97, 0.91}

\usepackage[frozencache]{minted}
\usepackage[minted, skins, breakable]{tcolorbox}

\usepackage{etoolbox}

\tcbset{
  colback=cosmiclatte,
	colframe=black!100,
  boxrule=0.5pt,
  top=0.2mm,
  bottom=0.2mm,
  left=1mm,
  listing only,
  minted options={
    tabsize=2,
  }
}
\setminted[]{
  fontsize=\footnotesize,
    tabsize=2,
	style=vs,
}
\renewcommand{\theFancyVerbLine}{\rmfamily\textcolor{cosmiclatte!50!black}{\tiny\arabic{FancyVerbLine}}} 
\newtcbinputlisting{externallisting}[3]{listing file={#2#3}, minted options={label=\textattachfile{#2#3}{#3}}, minted language={#1}}
\newtcblisting{internallisting}[1]{minted language=#1}
\newtcblisting{cppcode}{minted language=c++}

\definecolor{mygreen}{rgb}{0,0.8,0.6}
\definecolor{light-gray}{gray}{0.95} 
\definecolor{dkgreen}{rgb}{0,0.6,0}
\definecolor{gray}{rgb}{0.5,0.5,0.5}
\definecolor{mauve}{rgb}{0.58,0,0.82}
\definecolor{codegreen}{rgb}{0,0.6,0}
\definecolor{codegray}{rgb}{0.5,0.5,0.5}
\definecolor{codepurple}{rgb}{0.58,0,0.82}
\definecolor{backcolour}{rgb}{0.95,0.95,0.92}
\definecolor{orange}{RGB}{255,127,0}

\title{
Object-oriented design for
\\
massively parallel computing
}

\author{
	Edward Givelberg
}

\date{\today} 

\begin{document}

\maketitle 



\begin{abstract}
We define an abstract framework for object-oriented programming
and
show that object-oriented languages,
such as C++, can be interpreted
as parallel programming languages.
Parallel C++ code is typically more than ten times shorter than
the equivalent C++ code with MPI.
The large reduction in the number of lines of code in parallel C++
is primarily due to the fact that coordination of concurrency,
and the communications instructions,
including packing and unpacking of messages,
are automatically generated in the implementation of object
operations.
We implemented a prototype of a compiler and a runtime system
for parallel C++ and used them to create complex data-intensive
and HPC applications.
These results indicate that adoption of the parallel object-oriented
framework has the potential to drastically reduce the cost
of parallel programming.
We also show that
standard sequential object-oriented
programs can be ported to parallel architecture,
parallelized automatically, and potentially sped up.
The parallel object-oriented framework enables
an implementation of a compiler with a dedicated backend
for the interconnect fabric,
which
exposes the network hardware features directly
to the application.
We discuss the potential implications for computer architecture.
\end{abstract}


\section{Introduction}
\label{sec:introduction}

In 2004 the speed of individual processors reached a peak
and parallel computing
became a necessity.
It is now possible to build processors with thousands of cores,
yet
computer architecture is still based on von Neumann's design,
with operating systems incorporating only a limited amount of
parallelism, which is based on shared memory.
Sequential object-oriented languages,
such as C++, Java and Python,
remain the mainstream programming languages,
and
the most widely used model for parallel computation
is threads
(sequential processes that share memory).
The problems with threads are 
well known
(see, for example, 
\cite{Lee:2006:PT:1137232.1137289}).
The technology behind shared memory has grown in complexity,
with a a multi-level cache hierarchy consuming a lot of
chip area and energy.
\cite{Hennessy:2017:CAS:3207796}.
More generally, shared resources are potential bottlenecks
in a massively parallel computation,
and a distributed model is needed.

Object-oriented programming was originally 
inspired by the idea of imitating the real world
by organizing a computation
as a collection of separate (distributed) objects,
but it has been practiced primarily 
as a sequential programming paradigm.
In this paper we define an abstract framework for parallel 
object-oriented
computing and show that it is ideally suited for developing
massively parallel computer architecture.
The problem of
combining object-oriented design with 
parallel computing
has proved to be difficult.
The research literature is very extensive and the survey
articles
\cite{PhilippsenSurvey}
and
\cite{Boer:2017:SAO:3145473.3122848}
can be used as starting points to exploring the multitude
of approaches that have been proposed.
Perhaps the most difficult aspect of 
the problem
is the coordination
of concurrency
(see section 4 in \cite{PhilippsenSurvey}).
In practice,
coordination mechanisms require many lines of code and
place the burden of managing complex interactions on the programmer.
In a massively parallel computation
concurrency coordination cannot be micro-managed by the programmer,
and
we solve this problem by presenting
a high-level programming framework, where coordination takes place
automatically.
We implemented a prototype of a compiler and a runtime system,
and using these tools we were able to create data-intensive 
and HPC applications much faster (perhaps 10 times faster)
than we could do otherwise.
Our work contains several novel ideas which
combine to produce very powerful results:

\textbullet\ 
{\bf
We define an object as an abstract computer (a virtual machine).
}
An application
is
a collection of virtual processors
that is mapped onto a collection of physical processors,
in order 
to perform a given computational task.
Object-oriented
software architecture is a framework for classification
and construction of virtual processors,
and it can be used to design a compatible hardware architecture.

\textbullet\ 
{\bf
Automatic coordination of concurrency
as a result of causal relationships between objects.
}
Objects are autonomous entities.
Coordination results from
compiler-enforced
causal relationships between objects 
(i.e. an object may not use 
the results of a method execution before its completion).

\textbullet\ 
{\bf
Mainstream object-oriented languages can be interpreted as parallel 
programming languages.
}
The existing sequential standard can be extended to incorporate
parallel interpretation,
which
can be made available to the programmers
simply by providing a command-line compilation flag.

\textbullet\ 
{\bf
Sequential object-oriented code
can be parallelized either automatically, 
or with a small programming effort.
}
Standard sequential object-oriented code can be ported 
to run on parallel hardware.
We describe a parallelization technique which can increase
the efficiency of the serial code.


\textbullet\ 
{\bf
Parallel object-oriented code is 
high-level code.
}
Sequential code that uses a communications library
is exactly analogous to assembly code 
for a high-level programming language.
{\bf
We show that
parallel C++ code is
at least 10 times
shorter than the equivalent C++ code with MPI.
}
The large reduction in the number of lines of code in parallel C++
is primarily due to the fact that
synchronization and
communications instructions,
including packing and unpacking of messages,
are automatically generated in the implementation of object
operations.

\textbullet\ 
{\bf
We computed a 64 TB 3D Fourier tranform on a small cluster
to validate our parallel object-oriented
approach.
}
We used this challenging data-intensive problem as a benchmark.
A successful computation
depends on
good utilization of all of the system's components:
CPUs, network bandwidth
and the total available disk throughput.
The implementation 
requires extensive amount of coding using conventional methods.
We used 
about
500 lines of parallel C++ to solve the problem.

\textbullet\ 
{\bf
All network communications are compiler-generated instructions
implementing object operations.
}
This is the crucial feature of the object-oriented architecture.
It makes possible an implementation of 
a dedicated compiler backend for the interconnect fabric.


We define an abstract framework
for parallel object-oriented computing
in section \ref{sec:framework}.
Throughout this paper we use C++, but our results apply
to object-oriented languages in general.
In section \ref{sec:language} we show that
C++
can be interpreted as a parallel programming language
within the framework
of section \ref{sec:model}
without 
any change to the language syntax.
%
The implementation of the Fourier transform
(section \ref{sec:fft}),
and the
examples
in appendix \ref{sec:examples}, demonstrate that parallel C++
is a powerful and intuitive language.
In section
\ref{sec:compilerarch}
we outline the software
architecture for 
a parallel object-oriented compiler
and runtime system.
We describe our
prototype compiler for parallel C++ in the appendix
\ref{sec:prototype}.
In conclusion, we discuss some implications of parallel 
object-oriented
design for computer architecture.

%
%



\section{An abstract framework for object-oriented computing}
\label{sec:framework}

\subsection{What is an object?}

There is no consensus about the meaning of {\em object} 
in object-oriented programming.
The number of relevant abstractions
(remote objects,
fault-tolerant objects,
multicast objects,
tuple spaces,
etc.)
is too large to survey here.
We mention, but a few examples, which indicate the difficulty.
The wikipedia entry for object
\cite{WikipediaObject}
describes it as follows:
``an object can be a variable, a data structure, 
a function, or a method, and as such, 
is a value in memory referenced by an identifier''.
The C++ standard
defines an object as 
a region of storage
\cite{cppstandard},
while
in Python an object is described as an 
``abstraction for data''
\cite{PythonReference}.
On the other hand,
in
\cite{Ostrowski:2008:PLD:1428508.1428536} 
``live distributed objects'' are described as representing
running instances of distributed protocols,
but they have types and support composition,
much like ``ordinary'' objects.
Often, the definition of object is avoided
altogether,
even though it is of central importance in the design of
the programming framework.
We believe that
our definition of objects in
\ref{sec:model}
formalizes the ideas that originally led to the creation of
object-oriented programming.

\subsection{The Model}
\label{sec:model}

An {\em object} is 
an abstract autonomous parallel computing machine.
An {\em application}
is a collection of objects that perform a computation 
by executing methods on each other.
In an implementation,
an object is represented by an agent,
which is a collection of processes that receive
incoming method execution requests, 
execute them
and send the results back to the client objects.
An agent can process multiple method execution requests
simultaneously.
It can also represent several objects simultaneously,
effectively implementing
a {\em virtual host} where these objects live.

An object is accessed via a 
{\em pointer} (sometimes referred to as a remote pointer,
or a generalized pointer),
which contains the address of the virtual host representing the
object, as well as the address of the object 
within
the virtual host.

An application 
is started as a single object by the operating system,
which first creates
a virtual host and then constructs the application object on it.
The objects of the application may request the operating system to
create new virtual hosts and construct new objects on them.


\subsection{Related concepts}
\label{sec:processesactors}

A parallel computation is 
almost always
described as a collection
of concurrent coordinated processes,
however, in our opinion,
{\em
the concept of a computational process is not a suitable
abstraction for parallel programming.
}
In practice,
co-ordinating multiple concurrent processes in a computation
is a nearly impossible programming task,
even when the number of processes is small.
%
Objects, on the other hand,
are autonomous entities that coordinate naturally,
as we show in section
\ref{sec:causalasynchronous}.


The actor model
\cite{Hewitt:1973:UMA:1624775.1624804}
is an abstract model for distributed computing.
Actors perform computations and exchange messages,
and can be used for
distributed computing with objects.
A number of programming languages employ the actor model,
and many libraries and frameworks have been implemented
to permit actor-style programming in languages 
that don't have actors built-in
\cite{WikipediaActorModel}.
The main drawback of the actor model is that, like processes,
actors require coordination.
%
Furthermore,
the object-oriented model we introduced above
represents a higher level of abstraction.
The network, the computational processes and the messages 
are not included in the model,
and actors can be viewed as an implementation mechanism for objects.
In section \ref{sec:language}
we show that
in object-oriented languages
remote objects 
can be
constructed naturally,
using the existing language syntax.
This makes actor-model libraries 
which provide language bindings
redundant.





\section{Language interpretation}
\label{sec:language}

\subsection
{
Remote objects
}
\label{sec:remoteobjects}

\begin{figure}[H]
\centering
\begin{cppcode}
Host * host = new Host("machine1");
// C++ "placement new":
Object * object = new(host) Object(parameters);
result = object->ExecuteMethod(some, parameters);
\end{cppcode}
\caption{
Construction of an object on a remote hosts.
}
\label{fig:remote_ptr}
\end{figure}
The code in
Figure \ref{fig:remote_ptr} 
is compliant with the standard C++ syntax.
It creates a virtual host, constructs an object on it
and executes a remote method on that object.
We interpret all pointers as generalized pointers.
The virtual host object is provided by the operating system,
and is associated with a physical device.
In a massively parallel computation the construction of virtual hosts
and remote placement of objects will be done implicitly 
by the operating system.
In the execution of the remote method
by-value
parameters are serialized and sent over the network
to the remote host.
Once the remote execution completes, the result is sent back.
The treatment of by-reference parameters is more complicated:
the simplest solution is to serialize the parameter,
send it to the remote host and, upon completion of the method
execution, to serialize it and to send it back.
When the parameter is a complex object, and the changes made
by method execution are relatively small, there may be a more
efficient method to update the original parameter object.



\subsection{Distributed sequential execution}

With the introduction of remote objects
it is possible to execute an object-oriented program
on a distributed hardware platform in a manner
which is consistent with the standard sequential interpretation.
In {\em distributed sequential execution}
whenever an object executes a method on another (remote) object,
it waits for the completion of this operation
before executing the next statement.
While it is possible for several objects to simultaneously execute 
methods on a given object, this will never happen 
if the application is started as a single object.
In this case no parallel computation takes place,
but on some systems distributed sequential execution may offer 
advantages to single-processor sequential execution.
The additional cost of communication may be offset, for example, with 
faster execution 
or a lower energy cost on the dedicated remote hardware.
Furthermore, an improved overall system utilization can be achieved
if several parallel applications share the system's resources
(processors, interconnect, etc.).

\subsection{Causal asynchronous execution and coordination}
\label{sec:causalasynchronous}


We assume that an object is endowed with minimal intelligence
to enforce causality, i.e. to avoid using results of remote
execution before they become available.
The object can proceed with 
computation immediately
after initiating remote method execution,
and stop to wait for its completion only when its
results are needed.
We call this {\em causal asynchronous} execution.
This combination of
asynchronous communication 
\cite{DBLP:journals/sosym/JohnsenO07}
with the implicit future mechanism
\cite{Baker:1977:IGC:872734.806932}
is a natural consequence of the view that objects are
autonomous entities.
It enables the obejcts to operate in parallel
and does not require coordination by the programmer.
A simple example is shown in Figure \ref{fig:causal}.
%

\begin{figure}[H]
\centering
\begin{cppcode}
bool completed =
	remote_object->ExecuteMethod();
// do something while 
// the method is being executed
SomeComputation(); 
// wait for (remote) method completion
if (completed)
{
	// method execution has completed
	AnotherComputation();
}
\end{cppcode}
\caption{
Causal asynchronous execution and coordination.
The purpose of the {\tt if} statement is
to suspend the execution of {\tt AnotherComputation}
until the value of the variable
{\tt completed} is set by the remote method.
}
\label{fig:causal}
\end{figure}
\begin{figure}[H]
\vskip 0pt
\centering
\begin{cppcode}
void function()
{
	SomeComputation();
	{
		special_object->compute();
		for (int i = 0;  i < N;  i ++)
			object[i]->computation();
	}
	AnotherComputation();
}
\end{cppcode}
\caption{Nested compound statement.
The {\tt N + 1} statements
in the nested compound statement are executed in parallel
after {\tt SomeComputation} has completed.
{\tt AnotherComputation} is executed after the execution of
all of these 
{\tt N + 1} statements has completed.}
\label{fig:nestedCompound}
\end{figure}


Despite its simplicity, 
{\em parallel C++},
i.e. C++ with causal asynchronous interpretation,
has great expressive power
and is sufficiently rich to implement the most complex parallel
computations.
The programmer constructs a parallel computation by
coordinating high-level interactions of objects,
while
the low-level coordination and 
the underlying
network communications are generated by the compiler.
Computation and communication overlap naturally,
as in the example in Figure \ref{fig:causal},
and
large, complex objects can be 
sent over the network
as parameters of remote object methods.
This is ideally suited to utilizing high network bandwidth
and avoiding the latency penalty incurred by small messages.
%
In section
\ref{sec:parallelism}.
we introduce
additional mechanisms
for fine-grained control of parallelism.

\subsection {Automatic code parallelization}
\label{sec:parallelization}

The remote placement of objects can be applied to
all programs running on the system,
which can now be executed in a distributed sequential manner.
%
In order to parallelize a given program
remote object placement must be combined with causal asynchronous
code execution.
In many cases
control flow graph analysis
can be used to parallelize the code at compile time
(see examples in section
\ref{sec:examples}).
In addition to compile time analysis,
an object can avoid causality violation at runtime
by simply never using results of remote operations 
before they become available.
Such design does not prevent potential deadlocks.
Furthermore,
remote pointers can be abused, so
ultimately the programmer is responsible for
the causality correctness of the code.
Nevertheless,
we can expect any sufficiently complex sequential program
to contain code sections that parallelize automatically.


Automatic parallelization is a difficult and active area of
research.
Much of this work has focused on parallelizing loop execution.
Task parallelization typically requires the programmer to
use special language constructs to mark the sequential code,
which the compiler can then analyze for parallelization.
The {\em object-level parallelization} we introduced here
requires no new syntax, and,
arguably, well-structured serial object-oriented code can be
parallelized either automatically, 
or with minimal programming effort.

\subsection
{Detailed control of parallelism}
\label{sec:parallelism}


We now extend causal asynchronous execution to enable
a more fine-grained control of parallelism.
The interpretation defined in this section is one of many possible
parallel interpretations, and it is motivated by the examples
in section
\ref{sec:examples}.
In this model compound statements and iteration statements are
also executed asynchronously, and subject to causality.
In addition, we define the nested compound statement to be
{\em a barrier statement}.
This means that prior to its execution
the preceding statements in the parent compound statement
must finish execution, 
and its own execution 
must finish before the execution of the following statement starts
(see example in Figure
\ref{fig:nestedCompound}).
%
%
Similarly,
we require that
when the barrier statement is the first statement in
the iteration, 
the preceding iteration must finish before
the barrier statement is executed.
These definitions allow the programmer to describe
parallelism
in detail, 
as shown by the examples in
Figure \ref{fig:iterations}.


\begin{figure*}
\hrule
\vspace{2pt}
\begin{subfigure}[b]{1.0\columnwidth}
\begin{cppcode}
for (int i = 0;  i < N;  i ++)
{
	objectA[i]->computation();
	objectB[i]->computation();
}
\end{cppcode}
\caption{
\label{fig:iterations:parallel}
parallel
}
\end{subfigure}
\hfill
\begin{subfigure}[b]{1.0\columnwidth}
\begin{cppcode}
for (int i = 0;  i < N;  i ++)
{{
	objectA[i]->computation();
	objectB[i]->computation();
}}
\end{cppcode}
\caption{
\label{fig:iterations:seqiter}
sequential iterations
}
\end{subfigure}
\hfill
\begin{subfigure}[b]{1.0\columnwidth}
\begin{cppcode}
for (int i = 0;  i < N;  i ++)
{
	objectA[i]->computation();
	{
		objectB[i]->computation();
	}
}
\end{cppcode}
\caption{
\label{fig:iterations:pariter}
parallel iterations
}
\end{subfigure}
\hfill
\begin{subfigure}[b]{1.0\columnwidth}
\begin{cppcode}
for (int i = 0;  i < N;  i ++)
{
	{
		objectA[i]->computation();
	}
	objectB[i]->computation();
}
\end{cppcode}
\caption{
\label{fig:iterations:sequential}
sequential
}
\end{subfigure}
\vspace{1pt}
\caption{
\label{fig:iterations}
Iteration statements examples.
\\
\textbf { (\ref{fig:iterations:parallel}): }
potentially all 2N statements are executed in parallel.
\\
%
\textbf { (\ref{fig:iterations:seqiter}): }
all iterations are sequential,
but each iteration has two potentially parallel statements.
\\
%
\textbf { (\ref{fig:iterations:pariter}): }
potentially N iterations are executed
in parallel, but each iteration has 2 statements that are
executed sequentially.
\\
%
\textbf { (\ref{fig:iterations:sequential}): }
all 2N computations are executed sequentially.
}
\hrule
\end{figure*}
%




\section{Parallel C++ examples}
\label{sec:examples}

We present examples illustrating the expressive power of
parallel C++.

\subsection{Array objects}
\begin{figure}[H]
\vspace{-14pt}
\centering
\begin{cppcode}
double * a = new (remote_host) double([1024];
a[2] = 22.22  + x;
double z = a[24] + 3.1;
\end{cppcode}
\caption{Example: Array objects.}
\label{fig:arrays}
\end{figure}
The syntax of array operations 
applies naturally to remote pointers.
The array in the 
example in Figure \ref{fig:arrays}
is allocated on a remote host, and the 
array operations
require sending the values of {\tt x} and {\tt a[24]}
over the network.

\subsection{MapReduce}
\label{sec:MapReduce}

\begin{figure}
\centering
\begin{cppcode}
int NumberOfWorkers = 44444444;
Worker * workers[NumberOfWorkers];
for (int i = 0;  i < NumberOfWorkers;  i ++)
    workers[i] = new (host[i]) Worker();
for (int i = 0;  i < NumberOfWorkers;  i ++)
    result[i] = workers[i]->compute(data[i]);    
double total = 0.0;
for (int i = 0;  i < NumberOfWorkers;  i ++)
    total += result[i];
\end{cppcode}
\caption{Example: MapReduce.
The {\tt workers} array is assigned in parallel,
with each worker being constructed on its virtual host.
The {\tt compute} methods are also executed in parallel.
We rely on the compiler to enforce causality in the
execution of the reduction loop.
It starts executing only after {\tt result[0]}
becomes available, and it executes sequentially.
}
\label{fig:MapReduce}
\end{figure}

A basic example of 
MapReduce functionality can be implemented with
only a few lines of parallel C++ code, as shown in 
Figure \ref{fig:MapReduce}.
%
The master process allocates workers on remote hosts,
initiates a method execution on each worker
and sums up the result.
If the {\tt data[i]} object is not located on {\tt host[i]},
it will be copied there over the network.
This code is shorter and easier to write than the code that
uses Google's library.
Moreover, as we show in section
\ref{sec:compilerarch},
the parallel C++ compiler may be able to
generate more efficient code
by optimizing network operations.


%

\subsection{Breadth-First Search on a large graph}
\label{sec:bfs}

\begin{figure}
\centering
\begin{cppcode}
void Graph::BuildTree(VertexId root_id)
{
	int root_owner = VertexOwner(root_id);
	if (this->id() == root_owner)
		frontier.push_back(v[root_id]);
	EdgeList * E = new EdgeList[N];
	bool finished = false;
	while (!finished)
	{
		SortFrontierEdges(E);
		{
			// remote, asynchronous, in parallel
			for (int i = 0;  i < N;  i ++)
				graph[i]->SetParents(E[i]);
		}
		// finish BFS when all frontiers are empty
		finished = true;
		for (int i = 0;  i < N;  i ++)
			finished &= graph[i]->isEmptyFrontier();
	}
}
\end{cppcode}
\caption{
Building the BFS tree.
}
\label{fig:bfs}
\end{figure}
Distributed BFS on a large graph
is a standard benchmark problem
\cite{graph500}.
We implemented a straightforward algorithm in C++/MPI
using over 2000 lines of code,
and in parallel C++ with less than 200 lines of code.

The graph data is divided into N objects,
each containing an array of vertices
with a list of edges for each vertex.
We create N virtual hosts, one for each available processor,
and allocate a graph object on each host.
The main object initiates the BFS by invoking the
{\tt BuildTree} method on each graph object
(see Figure \ref{fig:bfs}).
The computation proceeds with several (typically less than 15)
synchronized iterations.
Each graph object keeps track of the local frontier,
which is a set of vertices on the current boundary
that have not been visited yet.
The graph object that owns the root vertex initializes its
local frontier with the root vertex.
In each iteration the local frontier edges are sorted into
N lists, one for each graph object.
The vertex on the other end of each frontier edge becomes
the child in the tree, unless it was visited before.
The new frontier set consists of the new children.
%
To set the parent links and to update the frontier set
every graph object executes 
a method on every other graph object, 
sending it the corresponding list of edges. 
This is done in the {\tt SetParents} method,
whose parameter is a large object of type
{\tt EdgeList},
which is serialized and sent over the network.
The calls to {\tt SetParents} execute in parallel
after the completion of {\tt SortFrontierEdges}.

BFS iterations stop when all local frontiers are empty.
We used $N^2$ messages to set the values of the {\tt finished}
variables.
This could be more conveniently achieved using an
allreduce library function, like those implemented in MPI.
In parallel C++ such functions could be, for example,
implemented in the standard library using specialized
containers for collective operations.
Notice that the {\tt while} iterations are executed
sequentially because they causally depend on the value
of {\tt finished}.

\subsection{Computation of a Fourier transform on a 64 TB 3D array}
\label{sec:fft}

\begin{figure}[H]
\centering
\includegraphics
[width=90mm]
{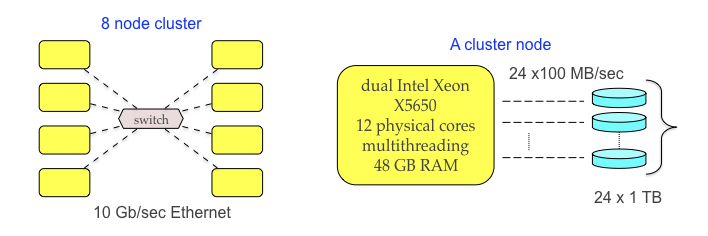}
\caption{
The 8-node cluster used in the 3D Fourier transform computation.
Every node of the cluster is connected in parallel 
to 24 one-terabyte hard drives, with 100 MB/sec read/write
throughput for each drive.
The CPUs, with 12 cores and 48GB of RAM each, are interconnected
with a 10 Gb/sec Ethernet.
}
\label{fig:cluster}
\end{figure}
%
%
%
\begin{figure}
\centering
\includegraphics
[width=70mm]
{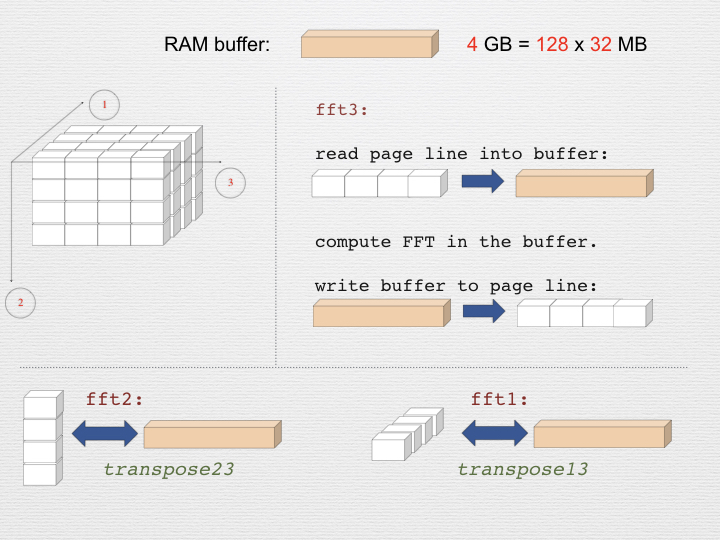}
\caption{
Fourier transform computation.
Lines of array pages are loaded into RAM buffers,
transformed and written back.
For the first and the second dimension additional
transpose operations are needed.
}
\label{fig:fft}
\end{figure}
%
%

%
\begin{figure*}
\begin{subfigure}[b]{1.0\columnwidth}
\begin{cppcode}
class Array 
{
public:
	Array(Domain * ArrayDomain,
						Domain * PageDomain);
	~Array();
	void allocate(int number_of_devices,
						Device * d);
	void FFT1(int number_of_cpus, Host ** cpus);
private:
	Domain * ArrayDomain;
	Domain * PageDomain;
	ArrayPage * *** page;	// 128^3 pointers
};
\end{cppcode}
\end{subfigure}
\hfill
\begin{subfigure}[b]{1.0\columnwidth}
\begin{cppcode}
void 
	Array::allocate(int number_of_devices,
						Device * d)
{
	for (int j1 = 0;  j1 < N1;  j1 ++) 
	for (int j2 = 0;  j2 < N2;  j2 ++) 
	for (int j3 = 0;  j3 < N3;  j3 ++) 
	{   
		// circulant
		int k = (j1 + j2 + j3) % number_of_devices;
		page[j1][j2][j3] =
			new(d[k]) ArrayPage(n1, n2, n3);
	}   
}
\end{cppcode}
\end{subfigure}
\caption{
\label{fig:ArrayClass}
The {\tt Array} class.
{\tt Domain} is a helper class describing 
3D subdomains of an array.
{\tt ArrayPage} is a small 3D array, which
implements local array operations,
such as 
{\tt transpose12} and {\tt transpose13} methods.
These operations are
needed in the Fourier transform computation.
Global array operations are implemented
using the local methods of {\tt ArrayPage}.
Array pages are allocated in circulant order.
The {\tt allocate} method 
constructs
${\tt N1} \times {\tt N2} \times {\tt N3}$
array pages
of size 
${\tt n1} \times {\tt n2} \times {\tt n3}$
on a list of virtual devices.
The dimensions are obtained from {\tt ArrayDomain}
and {\tt PageDomain}, and in our case are all equal 128.
}
\end{figure*}
%
%
We computed the Fourier transform of a 
64 TB array 
of $16384^3$ 
complex double precision numbers
on an 8-node cluster shown in Figure \ref{fig:cluster}.
%
%
%
%
%
The total computation time was approximately one day, and it could
be significantly improved with code optimization.
More importantly,
the hardware system could be redesigned
to achieve 
a better balance between the components.
Using more powerful hardware components a similar computation
can be carried out inexpensively with a 2 PB array on
a suitably configured small cluster.
We implemented the Fourier transform using approximately 15,000
lines of C++ code with MPI. 
The equivalent parallel C++ code is about 500 lines.

We used 4 of the cluster nodes to store the input array,
dividing it into $128^3$ pages of $128^3$ numbers each.
We used the other 4 nodes to run 16 processes 
of the Fourier transform,
4 processes per node.
An {\tt Array} object 
(see Figure \ref{fig:ArrayClass})
is logically a pointer:
it is a small object 
which is copied to all processes working on the array.
Array pages are allocated on 96 hard drives,
using virtual hosts for storing persistent objects,
which are implemented in the {\tt Device} class.
We ran two {\tt Device} agents on each CPU core,
each {\tt Device} 
using
a single hard drive.
In order to maximize 
the utilization of CPUs, network bandwidth
and the total available disk throughput,
array pages were allocated in circulant order
(see Figure \ref{fig:ArrayClass}).
%
%
%

\begin{figure}
\centering
\begin{cppcode}
void Array::FFT1(int number_of_cpus,
											Host ** cpu)
{
	int slab_width = N2 / number_of_cpus;
	SlabFFT1 ** slab_fft = 
		new SlabFFT1 * [number_of_cpus];

	for (int i = 0; 
				i < number_of_cpus;  i ++)
		slab_fft[i] = 
			new(cpu[i]) SlabFFT1(
							this, i * slab_width, 
							(i + 1) * slab_width);

	for (int i = 0;  i < n;  i ++)
		slab_fft[i]->ComputeTransform();
}
\end{cppcode}
\caption{
\label{fig:arrayfft}
Fourier transform of an array.
{\tt SlabFFT1} objects are constructed (in parallel)
on remote processors,
each {\tt SlabFFT1} is assigned a slab of the array.
The 16 {\tt SlabFFT1} objects compute the transforms
in parallel.
}
\end{figure}

The Fourier transform is computed by loading, 
and when necessary
transposing, lines of 128 pages into 4GB RAM buffers, performing
$128^2$ one-dimensional transforms in each buffer,
and writing the contents back to hard drives
(see Figure \ref{fig:fft}).
We illustrate the computation of the Fourier transform in
the first dimension.
%
%
%
\begin{figure}
\begin{subfigure}[t]{1.0\columnwidth}
\vskip 0pt
\begin{cppcode}
class SlabFFT1
{
public:
	SlabFFT1(Array * array, int N20, int N21);
	void ComputeTransform();
private:
	int N20, N21;	// slab indices
	Page * page_line, * next_page_line;	
	void ReadPageLine(
		ArrayPage * line, int i2, int i3
	);
	void WritePageLine(
		ArrayPage * line, int i2, int i3
	);
};
\end{cppcode}
\end{subfigure}
\hspace{1em}
\begin{subfigure}[t]{1.0\columnwidth}
\vskip 0pt
\begin{cppcode}
void SlabFFT1::ComputeTransform()
{
	ReadPageLine(page_line, N20, 0);
	for (int i2 = N20;	i2 < N21;	i2 ++)
	for (int i3 = 0;	i3 < N3;	i3 ++)
	{
		{
			int L2 = i2;
			int L3 = i3 + 1;
			if (L3 == N3)
				{  L3 = 0;    L2 ++;  }
			if (L2 != N21) 
				ReadPageLine(next_page_line, L2, L3);
			FFTW1(page_line);
		} // next_page_line has been read
		WritePageLine(page_line, i2, i3);
		page_line = next_page_line;
	}
}
\end{cppcode}
\end{subfigure}
\caption{
\label{fig:SlabFFT1}
Fourier transform of a slab.
\\
{\tt page\_line} and {\tt next\_page\_line} are 2 local RAM buffers,
4 GB each.
The iterations are sequential, and {\tt next\_page\_line}
is read while {\tt page\_line} is being transformed
using the
{\tt FFTW1} function,
which computes $128^2$ 1D FFTs using the FFTW library.
}
\end{figure}
\begin{figure*}
\begin{subfigure}[b]{1.0\columnwidth}
\begin{cppcode}
void SlabFFT1::ReadPageLine(
		ArrayPage * page_line, int i2, int i3)
{
	for (int i1 = 0;  i1 < N1;  i1 ++)
	{
		page[i1][i2][i3]->transpose13();
		page_line[i1] = *page[i1][i2][i3];
	}
}
\end{cppcode}
\end{subfigure}
\hspace{1em}
\begin{subfigure}[b]{1.0\columnwidth}
\begin{cppcode}
void SlabFFT1::ReadPageLine(
		ArrayPage * page_line, int i2, int i3)
{
	for (int i1 = 0;  i1 < N1;  i1 ++)
	{
		page_line[i1] = *page[i1][i2][i3];
		page_line[i1]->transpose13();
	}
}
\end{cppcode}
\end{subfigure}
\caption{
\label{fig:readpageline}
Two possible implementations of {\tt ReadPageLine}.
In our computation
we used the first implementation,
where the transpose is performed ``close to the data''
by the agent storing the page.
In the second implementation the transpose would be performed
by {\tt SlabFFT1} after it reads the page.
}
\end{figure*}
The processes computing the Fourier transform in the first
dimension are implemented in the {\tt SlabFFT1} class.
Each of the 16 {\tt SlabFFT1} objects was assigned 
an array slab of $128 \times 8 \times 128$ pages
to transform it line by line in $8 \times 128 = 1024$ iterations
(see Figure \ref{fig:arrayfft}).
The {\tt SlabFFT1} objects
are independent of each other
(see Figure
\ref{fig:SlabFFT1}),
but
they compete for service from the 96 hard drives,
and they share the network bandwidth.
The {\tt SlabFFT1} process
overlaps reading a page line with
{\tt FFTW1} function,
which computes $128^2$ 1D FFTs using the FFTW library
\cite{fftw}.
The 16 {\tt SlabFFT1} processes use the {\tt ReadPageLine} method
in parallel,
and each {\tt ReadPageLine} call reads 128 pages in parallel
from the hard drives storing the array,
and copies them over the network into the
RAM buffer of the {\tt SlabFFT1} object.
Figure
\ref{fig:readpageline}
shows two implementations of {\tt ReadPageLine},
demonstrating a very easy way to shift computation among processors.
We used the first implementation in our computation
in order to offload some of the work 
from the {\tt SlabFFT1} processes.
%
%
%
%
%
%
%
%
\begin{figure}
\centering
\begin{cppcode}
int main()
{
  int number_of_disks = 96;
  Device ** hdd = new Device *[number_of_disks];
  {
    for (int i = 0;  i < number_of_disks;  i ++)
      hdd[i] = new Device("hard drive i");
  }
  int n = 128;
  Domain page_domain(n, n, n);
  int N = n * n;
  Domain array_domain(N, N, N);
  Array * a = new Array(array_domain, page_domain);
  a->allocate(number_of_disks, hdd);
  int number_of_cpus = 16;
  Host ** cpu = new Host *[number_of_cpus];
  {
    for (int i = 0;  i < number_of_cpus;  i ++)
      cpu[i] = new Host("address of cpu i");
  }
  a->FFT1(number_of_cpus, cpu);
}
\end{cppcode}
\caption{
\label{fig:fftmain}
Fourier transform main
constructs in parallel 
96 virtual devices for array
storage, one on each hard drive of the 4 storage nodes.
The array object is created and array pages are allocated.
The 16 virtual hosts are created on the 4 computing nodes
after the page allocation has completed,
and the Fourier transform computation starts
after the construction of virtual hosts.
}
\end{figure}
%
%
%
%
The Fourier transform main 
(Figure \ref{fig:fftmain})
creates the array object and computes its transform.


\section{Compiler architecture}
\label{sec:compilerarch}

The object-oriented framework of section \ref{sec:model}
implicitly restricts network communications 
to implementation of object operations.
Object operations 
can be described
by an intermediate representation (IR) language.
We devised a rudimentary IR language
for our compiler prototype
(see the appendix \ref{sec:prototype}).
Here are three
examples of IR instructions: remote copy a block of memory,
initiate a method execution on an object, notify an agent
that a remote execution has completed.
IR code can be translated by the compiler into instructions
for the network hardware and for the CPUs.
It can also be used for compile-time analysis of 
network utilization,
as well as optimization of the system's performance as a whole.
It is therefore natural to implement a dedicated compiler backend
for the interconnect fabric.

The interconnect hardware instruction set is not restricted to
sending and receiving messages.
In the Mellanox InfiniBand, for example,
processing is done in network interface cards and network switches.
The following two examples illustrate the potential advantages of
compiler-generated networking instructions for
this network.

Applications must use large messages
to avoid the latency penalty and
to utilize
the network bandwidth.
As a result, a lot of code (and some processing power)
is devoted to packing and unpacking messages.
The User-mode Memory Registration (UMR) 
feature of Mellanox InfiniBand
can support MPI derived datatype communication,
which may reduce some of this overhead
\cite{DerivedDatatypeComm},
but it requires the programmer to duplicate
datatype definitions,
in order to inform the MPI library
about the datatypes used in the program.
In parallel C++ this information is available to the compiler,
which can generate the UMR instructions.

Another example of in-network processing is the 
Scalable Hierarchical Aggregation and Reduction Protocol
(SHARP) of Mellanox InfiniBand
\cite{Graham:2016:SHA:3018058.3018059}
that offloads the computation of collective operations, 
such as barrier and broadcast, to the switch network,
eliminating the need to send data multiple times between endpoints.
The SHARP hardware capabilities are currently accessed 
by the user only indirectly via a communications library, like MPI. 
However, the compiler 
is potentially capable of generating 
detailed and
efficient
routing and aggregation instructions
for very complex code.
Perhaps the simplest example is the following variant 
of the broadcast statement,
where a large number of objects 
{\tt a[i]} are located on some subset of the system's processors:
\begin{cppcode}
// a[i] are remote objects
for (int i = 0;  i < N;  i ++)
        a[i] = b;
\end{cppcode}

The last example suggests that the development of 
an optimizing compiler targeting network hardware
may lead to improved network hardware design.
In that respect, an especially
important example of a compilation target
is a many-core processor with a network-on-chip (NoC),
such as the Tile processor
\cite{tilera}.
Such processors can now be designed to
optimally execute IR code.


\label{sec:prototype}

\begin{figure}
\centering
\includegraphics
[width=85mm]
{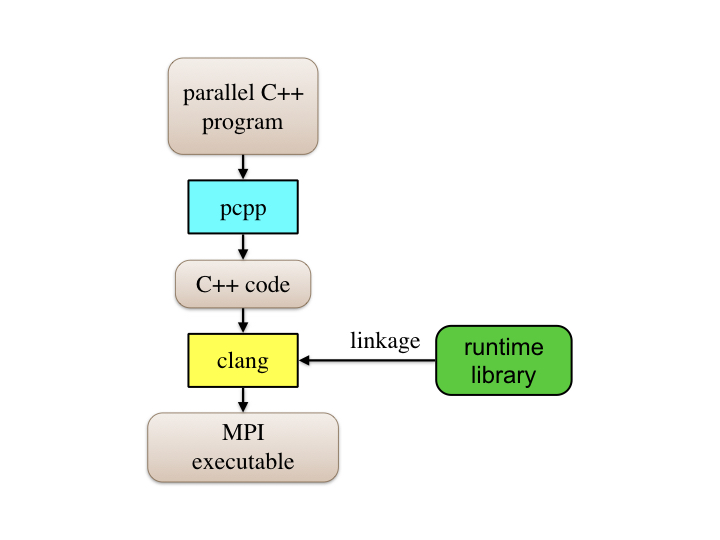}
\caption{
\label{fig:prototype}
Prototype compiler PCPP.
PCPP translates parallel C++ into C++ code,
which is compiled and linked against the runtime library to obtain
an MPI executable.
}
\end{figure}

We built a prototype compiler, called PCPP, and a runtime system
(see Figure \ref{fig:prototype}) for parallel C++.
The compiler translates parallel C++ into C++ code,
which is compiled and linked against the runtime library to obtain
an executable.
%
\subsection{The runtime library}
The runtime library implements virtual hosts
as agents that execute IR instructions.
All messages between agents are serialized IR instructions,
and for that purpose
the runtime library contains a simple serialization layer.
An agent is implemented as an MPI process with multiple threads:
a dispatcher thread and a pool of worker threads.
The dispatcher thread receives an
incoming message, unserializes it into an IR instruction
and assigns it to a worker thread for execution.
Each worker thread maintains a job queue of IR instructions,
however the pool of worker threads is not limited,
and can grow dynamically.
Every worker thread is either processing its job queue,
is suspended and waiting to be resumed, or is idle and
available to work.
An execution of an IR instruction typically involves
execution of the application's code
and may result in new IR instructions being sent over the network.
We used one dedicated worker thread in every agent
to serialize and send
IR instructions to their destination agents.

We used a small number of basic MPI commands to implement
a transport library for agents' communications,
and to launch agents on remote hosts as MPI processes.
All of the MPI functionality used in the prototype 
is encapsulated in the transport library and can be easily replaced.


\subsection{The PCPP compiler}

PCPP is source-to-source translation tool which works with a subset
of the C++ grammar. It is built using the Clang library tools.
(Clang is the front end of the LLVM
compiler
software
\cite{Lattner:2004:LCF:977395.977673}.)
PCPP transforms the main of the input program into a stand-alone 
class, the application's main class.
It generates a new main program which 
initializes the runtime system, 
constructs a virtual host and
constructs the application's main object on it.
Next, the new main reverses these actions,
destroying the application's main object on the virtual host,
destroying the virtual host and shutting down the runtime system.

PCPP translates all pointers to remote pointer objects.
For every class of the application PCPP generates
IR instructions for its object operations
(constructors, destructors and methods).
Additionally, PCPP replaces calls to object operations
with code that serializes the parameters and sends them with
the corresponding instruction to the destination agent.
For example,
when a constructor is invoked,
one of the serialized parameters is a remote pointer containing
the address of the result variable,
which is a remote pointer variable that should be assigned
with the result of the constructor.
The PCPP-generated IR instruction is a serializable class, 
derived from
the base instruction class defined in the runtime library.
When this instruction is received by the destination agent,
it is unserialized and its {\tt execute} method is invoked.
This method constructs a local object using the 
unserialized
parameters and generates an IR instruction to copy the object
pointer to the result variable on the source agent.

For causality enforcement we implemented a simple {\tt guard}
object, based on
the {\tt condition\_variable} of the C++11 standard library.
PCPP generates a guard object for every output variable of
a remote operation.
A {\tt wait} method on the guard object suspends the executing thread
until a {\tt release} method is called on the same guard object
by another thread.
A remote pointer to this guard object is 
sent to the destination agent.
When the destination agent completes the operation
it sends an IR instruction to the source agent to release the guard.
The {\tt wait} call is inserted in the application code
just before the value of the output
variable is used.

\section{Conclusion and Future Work
}
\label{sec:architecture}

We have defined a framework for object-oriented computing
and have shown that object-oriented languages can be interpreted
in this framework as parallel programming languages.
Parallel C++ is a very powerful language.
We have completed a basic working prototype
of the compiler and the runtime system,
which we used to implement complex applications
whose implementation would be 
very expensive using conventional techniques.
We have shown that standard sequential C++ programs
can be ported to 
parallel hardware, 
parallelized automatically, and potentially sped up.
A large
amount of work
is required to move from
the current prototype stage to a fully functional product,
but our results indicate that
the adoption of parallel C++ has the potential to
drastically reduce the cost of parallel programming.

We are very excited about the implications
of the object-oriented framework for computer architecture.
Processors with a large number of cores and a network on chip 
(NoC) are very energy efficient
\cite{Francesquini:2015:EEP:2780684.2780859},
but are very difficult to program
\cite{Mattson:2008:PIN:1413370.1413409}.
We propose
the object-oriented framework as a possible solution.
The processing cores can be optimized to implement virtual hosts.
To increase energy efficiency,
the system can be designed with a variety of 
specialized processing
cores, with the compiler mapping objects to appropriate cores.
Because each object has its own address space,
the need for unlimited address space is relaxed,
and a distributed memory architecture can be designed to replace
the complex single shared memory.
The parallel object-oriented compiler is probably the
most important component in the construction of
an operating system for a multi-processor computer.
With a dedicated compiler backend for the interconnect fabric,
the network is no longer a passive collection of data pipes.
The compiler exposes the hardware features 
of the network architecture directly to the applications.
The compiler may be able to generate more efficient code,
analyse network traffic and provide the operating system with
the means to control the congestion on the NoC.
%
In present systems such application awareness is very difficult
to implement
\cite{Nychis:2010:NGO:1868447.1868459}.

\newpage

%

\newpage
\newpage


\bibliography{paper}
\bibliographystyle{plain}


\end{document}

%% file: structure.tex
\usepackage[english]{babel} 

\usepackage{microtype} 

\usepackage{amsmath,amsfonts,amsthm} 

\usepackage[svgnames]{xcolor} 

\usepackage[hang, small, labelfont=bf, up, textfont=it]{caption} 

\usepackage{booktabs} 

\usepackage{lastpage} 

\usepackage{graphicx} 

\usepackage{enumitem} 
\setlist{noitemsep} 

\usepackage{sectsty} 
\allsectionsfont{\usefont{OT1}{phv}{b}{n}} 

\usepackage{geometry} 

\usepackage[T1]{fontenc} 
\usepackage[utf8]{inputenc} 

\usepackage{XCharter} 

\usepackage{fancyhdr} 

\fancypagestyle{firstpage}{ 
	\fancyhf{}
}

\usepackage{lettrine} 
\usepackage{fix-cm}	

\usepackage{xstring} 

